# Ultrahigh thermal conductivity in hexagonal BC$_6$N- An efficient material for nanoscale thermal management- A first principles study

Rajmohan Muthaiah, Jivtesh Garg

School of Aerospace and Mechanical Engineering, University of Oklahoma, Norman, OK-73019, USA

**Abstract:** Engineering materials with high thermal conductivity are of fundamental interest for efficiently dissipating heat in micro/nanoelectronics. Using first principles computations we report an ultra-high thermal conductivity of 2090 Wm$^{-1}$K$^{-1}$ (1395 Wm$^{-1}$K$^{-1}$) for hexagonal pure (natural) BC$_6$N($h$-BC$_6$N). This value is among the highest thermal conductivities known after diamond and cubic boron arsenide. This ultra-high lattice thermal conductivity ($k$) is mainly attributed with high phonon group velocities of both acoustic and optical phonons arising from strong C-C and B-N bonds as well as the light atomic mass of the constituent elements such as boron (B), carbon (C) and nitrogen (N). We also report size dependent thermal conductivity of $h$-BC$_6$N nanostructures by including boundary scattering. At room temperature (300 K) and at nanoscale length (L) of 100 nm, a high $k$ value of 175 Wm$^{-1}$K$^{-1}$ is observed (higher than the bulk $k$ value of silicon). Optical phonons with large group velocities are mainly responsible for this high thermal conductivity in $h$-BC$_6$N nanostructures. High thermal conductivity of $h$-BC$_6$N makes it a candidate material for heat dissipation in micro/nano thermal management applications.

**Introduction:** Continued decrease in transistor length has led to the issue of hot spots[1] in electronic chips. Heat removal in micro/nanoelectronics is a fundamental issue that limits its performance and reliability[2, 3]. High thermal conductivity materials can play a strong role in improving heat dissipation. Cubic[4-6] and hexagonal diamond[7] (also known as lonsdaleite) has been reported as the most thermally conductive materials on earth due to the strong covalent bonds and light atomic mass of carbon atoms. Likewise, boron[8-16] based compounds were reported with high thermal conductivity in bulk and nanostructured materials due to high phonon group velocity arising from the light atomic mass. Recently, Sadeghi *et al*., reported a high thermal conductivity of 2073 Wm$^{-1}$K$^{-1}$ for the hexagonal BC$_2$N($h$-BC$_2$N)[8] with optical phonons has considerable phonon group velocities which are beneficial for the thermal transport in nanostructures. Likewise, Shafique *et al.,* reported a high thermal conductivity of 1275 Wm$^{-1}$K$^{-1}$ for the monolayer graphene like

$BC_2N$[12]. Mortazavi *et al.,* reported an ultra-high thermal conductivity 1710 $Wm^{-1}K^{-1}$ for the monolayer $BC_6N$[17] but its counterpart of bulk thermal conductivity is yet to be reported. In this work, we analyzed the thermal conductivity of bulk and nanostructured hexagonal $BC_6N$($h$-$BC_6N$) through first principles calculations. Our first principles calculations reveal an ultra-high bulk thermal conductivity ($k$) of 2090 $Wm^{-1}K^{-1}$(1395 $Wm^{-1}K^{-1}$) for the pure (natural) hexagonal $BC_6N$ which is among the highest thermal conductivity values ever reported (lower only to diamond and boron arsenide[13]). Likewise, at nanometer length scales such as 100 nm, computed room temperature thermal conductivity of 175 $Wm^{-1}K^{-1}$, indicating $h$-$BC_6N$ will be a candidate material for thermal management in nanoelectronics. We systematically investigated the elastic constants, mode-contribution thermal conductivity of transverse acoustic (TA), longitudinal acoustic (LA) and optical phonon modes, phonon group velocity, phonon scattering rates and phonon mean free paths. We noticed that, optical phonons with considerable phonon group velocities and phonon lifetimes, contribute significantly to overall thermal conductivity in $h$-$BC_6N$.

**Computational methods:** First principles computations were performed using local density approximations[18] with norm-conserving pseudopotentials within QUANTUM ESPRESSO[19] package. The geometry of the hexagonal $BC_6N$($h$-$BC_6N$) with 8 atoms unit cell is optimized until the forces on all atoms are less than $10^{-5}$ ev $Å^{-1}$ and the energy difference is converged to $10^{-15}$ Ry. A plane-wave cutoff energy of 80 Ry was used. Electronic calculations were performed using 12 x 12 x 4 Monkhorst-Pack[20] $k$-point mesh. Optimized $h$-$BC_6N$ structure with lattice constants of a=2.4802 Å and c/a=3.3438 is shown in Fig. 1. Elastic constants were computed using QUANTUM ESPRESSO thermo_pw package and Voigt-Reuss-Hill approximation[21] is used to calculate the bulk modulus(B), Young's Modulus(E) and shear modulus(G). Dynamical matrix and harmonic force constants were computed using 12 x 12 x 4 q-grid. 6 x 6 x 2 q-points were used to compute the anharmonic force constants using QUANTUM ESPRESSO D3Q[22-24] package. Acoustic sum rules were imposed on both harmonic and anharmonic interatomic force constants. Lattice thermal conductivity($k$) is calculated by solving phonon Boltzmann transport equation (PBTE)[22, 24, 25]. Thermal conductivity ($k$) in the single mode relaxation time (SMRT) approximation[26] (usually the first approximation in solution of PBTE) is given by,

$$k_\alpha = \frac{\hbar^2}{N\Omega k_b T^2} \sum_\lambda v_{\alpha\lambda}^2 \omega_\lambda^2 \bar{n}_\lambda (\bar{n}_\lambda + 1)\tau_\lambda \tag{1}$$

where, $\alpha, \hbar, N, \Omega, k_b, T$, are the cartesian direction, Planck constant, size of the q mesh, unit cell volume, Boltzmann constant, and absolute temperature respectively. $\lambda$ represents the vibrational mode ($qj$) ($q$ is the wave vector and $j$ represent phonon polarization). $\omega_\lambda, \bar{n}_\lambda$, and $v_{\alpha\lambda} (= \partial\omega_\lambda/\partial q)$ are the phonon frequency, equilibrium Bose-Einstein population and group velocity along cartesian direction $\alpha$, respectively of a phonon mode $\lambda$. $\omega_\lambda, \bar{n}_\lambda$, and $c_{\alpha\lambda}$ are derived from the knowledge of phonon dispersion computed using 2$^{nd}$ order IFCs. $\tau_\lambda$ is the phonon lifetime and is computed using the following equation,

$$\frac{1}{\tau_\lambda} = \pi \sum_{\lambda'\lambda''} |V_3(-\lambda, \lambda', \lambda'')|^2 \times [2(n_{\lambda'} - n_{\lambda''})\delta(\omega(\lambda) + \omega(\lambda') - \omega(\lambda'')) + (1 + n_{\lambda'} + n_{\lambda''})\delta(\omega(\lambda) - \omega(\lambda') - \omega(\lambda''))] \tag{2}$$

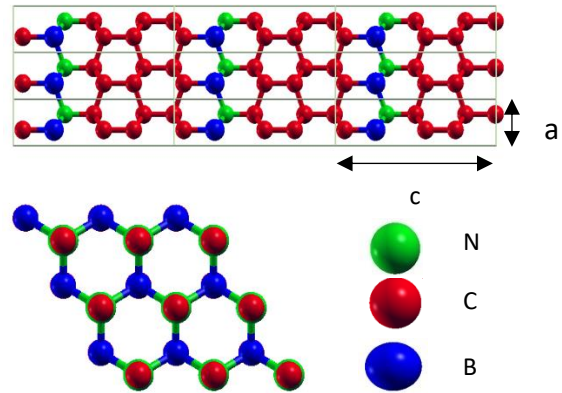

Figure 1: $h$-$BC_6N$ crystal structure with lattice parameters a= 2.4802 Å and c/a=3.3438. Blue, red and green represents the boron, carbon and nitrogen atoms respectively.

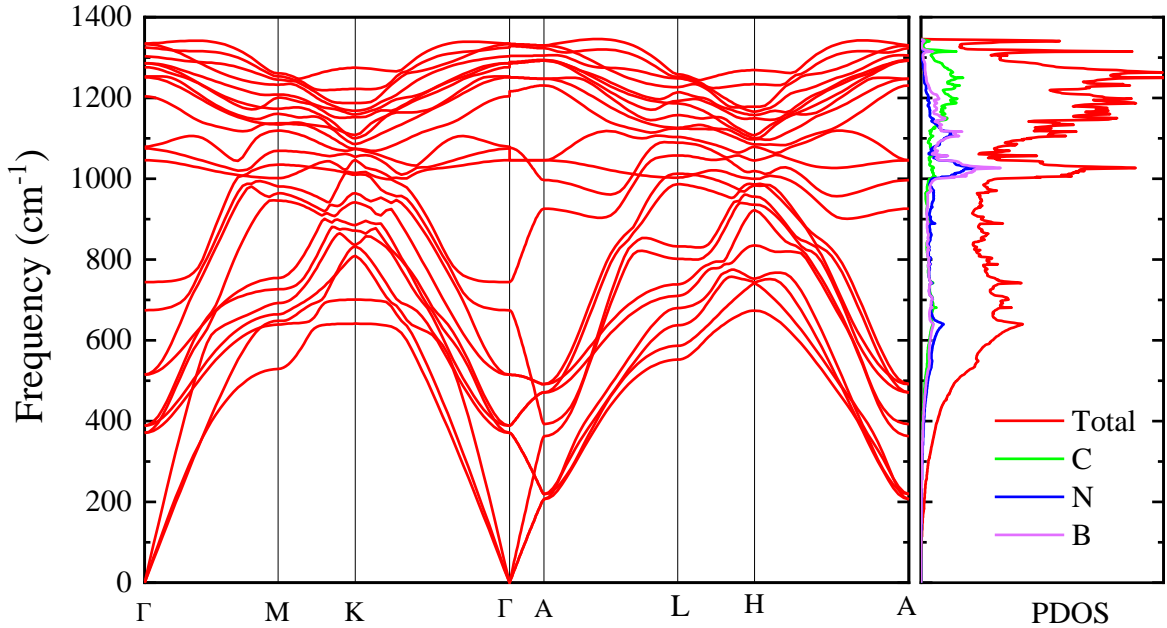

Figure 2: Phonon dispersion and phonon density of states for the hexagonal BC$_6$N with lattice constants a= 2.4802 Å and c/a=3.3438. Green, blue and purple color represents the contributions from carbon (C), nitrogen (N) and boron (B) respectively.

where, $\frac{1}{\tau_\lambda}$ is the anharmonic scattering rate due to phonon-phonon interactions and $V_3(-\lambda, \lambda', \lambda'')$ are the three-phonon coupling matrix elements computed using both harmonic and anharmonic interatomic force constants. Phonon linewidth and lattice thermal conductivity were calculated iteratively using QUANTUM ESPRESSO thermal2 code with 30 x 30 x 10 q -mesh and 0.05 cm$^{-1}$ smearing until the $\Delta k$ values are converged to 1.0e$^{-5}$ Wm$^{-1}$K$^{-1}$. Iterative results were converged after 6 iterations. Casimir scattering[27] is imposed for size dependence thermal conductivity calculations for the nanostructured $h$-BC$_6$N.

**Results and Discussion:**

Phonon dispersion and phonon density of states for the hexagonal BC$_6$N with its equilibrium lattice constants a= 2.4802 Å and c/a=3.3438 is shown in Fig 2. Positive phonons frequencies of all the phonon branches indicate the dynamical stability of $h$-BC$_6$N. To validate the mechanical stability of the system, we calculated elastic constants to check the Born stability criteria[28]. Elastic constants of $h$-BC$_6$N are listed in Table 1 and compared against $h$-diamond and $h$-BC$_2$N. The calculated elastic constants satisfied the Born stability criteria of $C_{66}=(C_{11}-C_{12})/2$, $C_{11} > C_{12}$, $C_{33}(C_{11}+C_{12}) > 2(C_{13})$, $C_{44} > 0$, $C_{66} > 0$, and hence the system is mechanically stable. Bulk modulus(B), Young modulus(E) and Shear modulus(G) based on Voigt-Reuss-Hill approximation are also listed in Table 1. The computed values are slightly lower than the hexagonal diamond. These elastic constants and positive frequencies in phonon dispersion indicate the mechanical and dynamical stability of $h$-BC$_6$N.

**Table 1: Elastic constants (in GPa) of $h$-BC$_6$N, $h$-Diamond and $h$-BC$_2$N.**

| Material | C11 | C33 | C44 | C66 | C12 | C13 | Bulk Modulus(B) | Young modulus(E) | Shear Modulus(G) |
|---|---|---|---|---|---|---|---|---|---|
| $h$-BC$_6$N | 1182.98 | 1298.11 | 438.90 | 537.20 | 108.58 | 20.33 | 440.00 | 1107.40 | 512.30 |
| $h$-Diamond | 1251.52 | 1367.74 | 483.00 | 579.40 | 92.61 | 20.00 | 450.74 | 1182.45 | 556.31 |
| $h$-BC$_2$N | 1091.10 | 1146.23 | 399.50 | 498.70 | 93.60 | 2.97 | 391.90 | 1007.40 | 470.05 |

Temperature dependence lattice thermal conductivity($k$) of the pure (solid lines) and naturally (dotted lines) occurring $h$-BC$_6$N, obtained by solving the phonon Boltzmann transport equation exactly, is shown along a-axis and c-axis in Fig 3a. At 300 K, thermal conductivity of pure and naturally occurring $h$-BC$_6$N is 2090 Wm$^{-1}$K$^{-1}$(1082 Wm$^{-1}$K$^{-1}$) and 1395 Wm$^{-1}$K$^{-1}$(832 Wm$^{-1}$K$^{-1}$) respectively along the a-axis (c-axis). $k$ of naturally occurring $h$-BC$_6$N is 33.25% lower than the pure $h$-BC$_6$N. Thermal conductivity for naturally occurring $h$-BC$_6$N was computed by introducing phonon scattering arising out of mass-disorder due to random distribution of isotopes of Boron, Carbon and Nitrogen throughout the crystal. The small mass variation in isotopes of B (atomic mass of 10.012 a.u with 19.9% concentration and atomic mass of 11.009 a.u with 80.1% concentration), C (atomic mass of 12 a.u with 98.93% concentration and 13.0033 a.u with 1.07% concentration) and N (atomic mass of 14.003 a.u with 99.636% concentration and 15.0 a.u with 1.07% concentration) atoms[29], induces moderate additional phonon scattering, causing only 33.25% decrease in thermal conductivity of naturally occurring $h$-BC$_6$N relative to the pure case. Lattice thermal conductivity using single mode relaxation approximation (SMA) is shown in supplementary Fig S1. At 300K, lattice thermal conductivity of $h$-BC$_6$N within the SMA is 1601 Wm$^{-1}$K$^{-1}$ (999 Wm$^{-1}$K$^{-1}$) along the a-axis (c-axis); this value is 23.3% (7.67%) lower than the iterative solution.

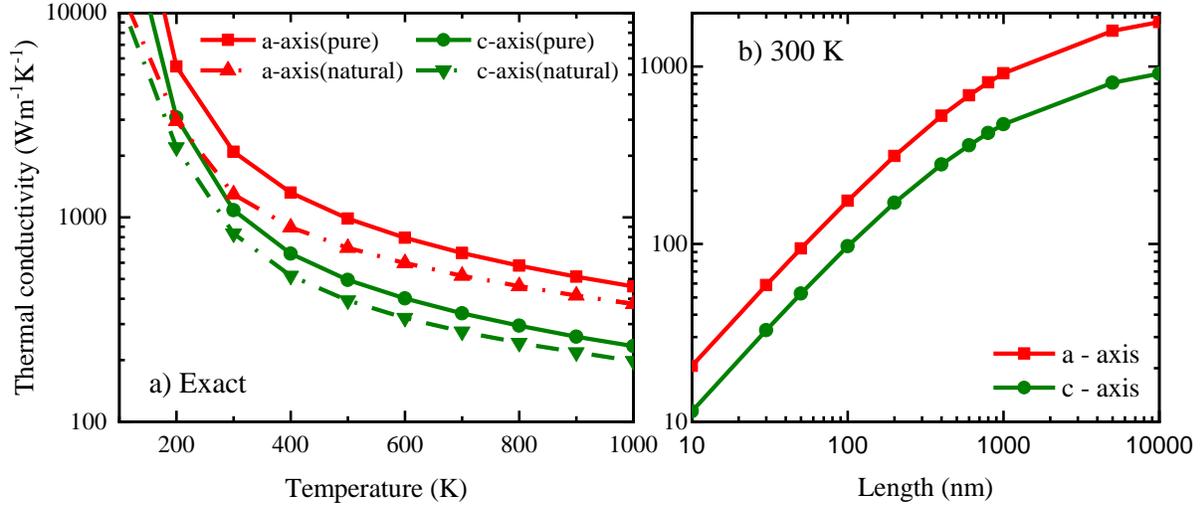

Figure 3a): Temperature dependent lattice thermal conductivity along a and c axis for pure and naturally occurring $h$-BC$_6$N. b) Room temperature size dependence thermal conductivity of pure $h$-BC$_6$N.

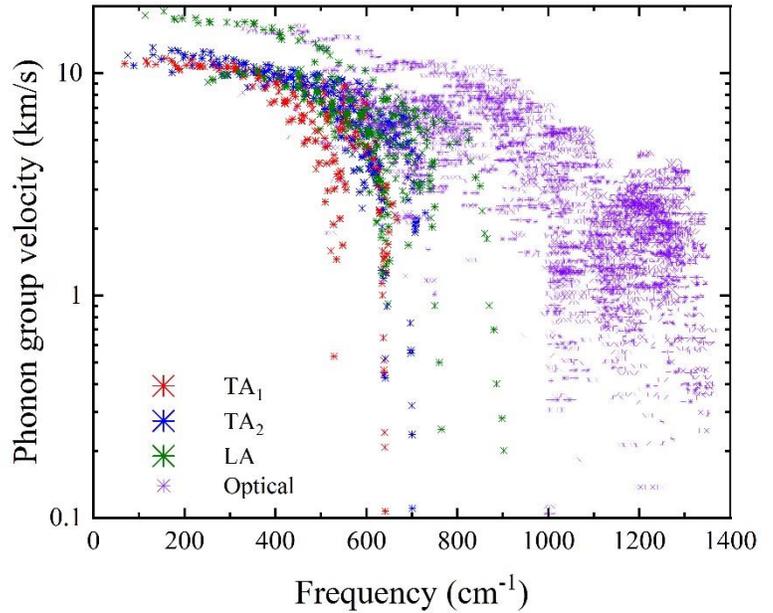

Figure 4: Phonon group velocity of TA$_1$, TA$_2$, LA and Optical phonon modes of $h$-BC$_6$N.

This ultra-high thermal conductivity of $h$-BC$_6$N is mainly attributed to high phonon frequencies ($\omega_\lambda$) and phonon group velocities ($= \partial\omega_\lambda/\partial q$) of both acoustic and optical phonons. These high frequencies are due to the strong C-C, B-C and B-N bonds and the light atomic mass of the constituent atoms B, C and N. These strong bonds are verified by the elastic constants presented in Table 1 with diamond and $h$-BC$_2$N. Bulk modulus and Young modulus for $h$-BC$_6$N is close to that of diamond. This is due to the strong covalent bond network through $sp^3$ hybridization of the atoms.

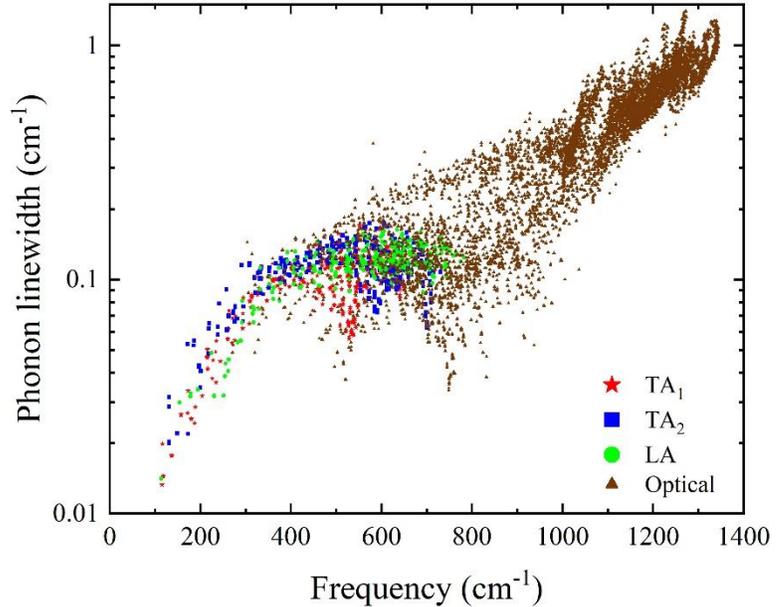

Figure 5: Phonon linewidths of TA$_1$, TA$_2$, LA and optical phonons for $h$-BC$_6$N. Red, Blue, Green and brown color represents the TA$_1$, TA$_2$, LA and optical phonons respectively.

To elucidate further, we analyzed the mode contribution thermal conductivity, phonon group velocities and phonon scattering rates of transverse acoustic (TA), longitudinal acoustic (LA) and optical phonons. At 300 K, we observed that, TA$_1$, TA$_2$, LA and optical phonons contribute 21.9% (16.62%), 18.11% (15.77%), 17.37% (18.87%) and 42.62% (48.74%) to overall thermal conductivity, respectively, along a-axis (c-axis). Typically, acoustic phonons are considered as major heat carrier phonons and optical phonons serve as a scattering channels for the acoustic phonons[30]. It is interesting to note that in BC$_6$N, optical phonons also contribute significantly to overall thermal conductivity. This is mainly due to the fact that optical phonons have considerable high phonon group velocities (Fig 4) and phonon lifetimes (inverse of the phonon linewidths as shown in Fig 5).

To explore the thermal transport in nanostructures, we have computed the size dependent thermal conductivity of $h$-BC$_6$N by introducing phonon boundary/Casimir[27] scattering. We have computed the length dependence of only pure $h$-BC$_6$N. Length dependent thermal conductivity($k$) of $h$-BC$_6$N between 10 nm and 10 um is shown in Fig 3b. For an example, at nanometer length scales of L = 100 nm, room temperature $k$ of $h$-BC$_6$N is 175 Wm$^{-1}$K$^{-1}$ along the a-axis which is significantly higher than the thermal conductivity of bulk silicon[31]. To understand this further, we computed the phonon mean free paths of TA$_1$, TA$_2$, LA and optical phonon modes as shown in Fig 6. We can observe that acoustic phonons have mean free path higher than 100 nm and will be scattered significantly for nanostructures smaller than 100 nm, due to boundary scattering. Optical phonon meanfreepaths, however, are in the range of nanometers, contributing to the high nanoscale thermal conductivity in $h$-BC$_6$N.

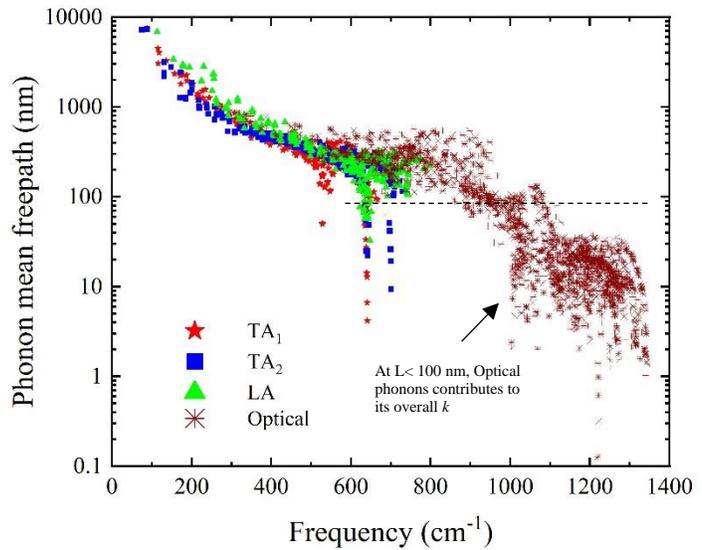

Figure 6: Phonon mean free path of TA$_1$, TA$_2$, LA and Optical phonons for $h$-BC$_6$N

**Conclusion:** In this work, by solving Boltzmann transport equation with first principles calculations, we report an ultra-high lattice thermal conductivity of 2090 Wm$^{-1}$K$^{-1}$(1082 Wm$^{-1}$K$^{-1}$) along a-axis(c-axis) for pure hexagonal BC$_6$N($h$-BC$_6$N) which is the 3$^{rd}$ highest reported thermal conductivity after diamond and cubic boron arsenide. This ultra-high thermal conductivity is mainly attributed to high phonon frequencies and phonon group velocities arising from the strong C-C, B-C and B-N bonds and the light atomic mass of the constituent atoms B, C and N. We also observed a significant phonon scattering due to isotopic disorder. Elastic constants show that $h$-BC$_6$N is ultrahard with elastic constants almost equal to diamond. We also computed the size dependent thermal conductivity of *h*-BC$_6$N between 10 nm and 10000 nm. At nanometer length scales of L=100 nm, a high room temperature thermal conductivity of 175 Wm$^{-1}$K$^{-1}$ was reported. This points to the promising nature of *h*-BC$_6$N as candidate for nanoscale thermal management applications.

**Conflicts of Interest**

There are no conflicts of interest to declare.

**Acknowledgements**

RM and JG acknowledge support from National Science Foundation CAREER award under Award No. #1847129. We also acknowledge OU Supercomputing Center for Education and Research (OSCER) for providing computing resources for this work.

**Data Availability:** The raw/processed data required to reproduce these findings can be shared upon request.

**Supplementary Information**

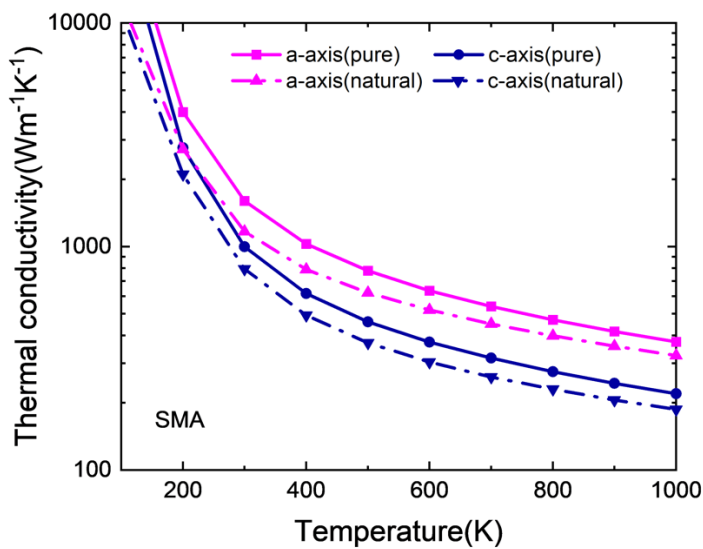

Figure S1: Temperature dependence lattice thermal conductivity of pure and naturally occurring h-BC6N along a and c axis.